\documentclass[showpacs]{revtex4}
\usepackage{fullpage}
\usepackage[dvips]{graphicx}
\input{epsf}
\usepackage{amsmath}
\usepackage{amssymb}
\usepackage[mathscr]{eucal}
\usepackage{bm}

\newcommand{\beq}{\begin{equation}}
\newcommand{\eeq}{\end{equation}}
\newcommand{\beqa}{\begin{eqnarray}}
\newcommand{\eeqa}{\end{eqnarray}}
\newcommand{\beqan}{\begin{eqnarray*}}
\newcommand{\eeqan}{\end{eqnarray*}}

\newcommand{\tr}[1]{{\rm tr} \left( #1 \right) }
\newcommand{\ket}[1]{| #1 \rangle}
\newcommand{\bra}[1]{\langle #1 |}

\newcommand{\llangle}{\langle\!\langle} 
\newcommand{\rrangle}{\rangle\!\rangle}

\newcommand{\proof}{\noindent {\bf Proof. }}
\newcommand{\qed}{\hfill $\Box$ \vskip 2ex}

\newtheorem{theorem}{Theorem}
\newtheorem{definition}{Definition}

\newtheorem{corollary}{Corollary}

\begin{document}

\title{Controllability properties for finite dimensional quantum Markovian master equations}
\author{Claudio Altafini}
\affiliation{SISSA-ISAS  \\
International School for Advanced Studies \\
via Beirut 2-4, 34014 Trieste, Italy }
\email{altafini@sissa.it}


\begin{abstract}
Various notions from geometric control theory are used to characterize the behavior of the Markovian master equation for N-level quantum mechanical systems driven by unitary control and to describe the structure of the sets of reachable states.
It is shown that the system can be accessible but neither small-time controllable nor controllable in finite time.
In particular, if the generators of quantum dynamical semigroups are unital, then the reachable sets admit easy characterizations as they monotonically grow in time.
The two level case is treated in detail.

\end{abstract}

\maketitle 


\section{Introduction}

The main question that we discuss in this work is the following: to which density operators can we drive the quantum Markovian master equation by means of coherent control?
This problem is of relevance whenever one is interested in quantum state manipulation in presence of nonunitary evolution, for example in the context of quantum information processing \cite{Bacon1,Lloyd2,Recht1} and of molecular control \cite{Tannor1}.
The ultimate goal is obviously to know when and how the state of a quantum mechanical system can be arbitrarily manipulated by means of unitary (reversible) control operations or at least to what extent this is possible. 

The viewpoint we take in this work is that of ``classical'' geometric  control theory which provides us the tools to mathematically formalize and answer the questions posed.
In classical control terms, the set of density operators to which we can steer the system is called the {\em reachable set} and the problem of arbitrary manipulability of the state can be formulated as a {\em controllability} problem.

The infinitesimal structure of the so-called quantum Markovian master equation, i.e. the ``axiomatic'' model for an open quantum system, is known since the works of Lindblad \cite{Lindblad1} and Gorini-Kossakowski-Sudarshan \cite{Gorini1} and it is a prerequisite for the utilization of the Lie algebraic controllability methods developed below.
We use the so-called {\em vector of coherences} formulation for the density matrix \cite{Alicki1}, i.e. the expectation values corresponding to a complete set of Hermitian operators, here the Gell-Mann matrices.
Such formulation allows to treat the master equation as a control system with affine vector fields or, geometrically, as a system living on a homogeneous space of a matrix Lie group and subordinated to an affine group action, plus constraints originating from the complete positivity of the quantum dynamical semigroup.
If we drop these constraints, the system falls into a class of systems whose controllability properties were studied in detail in the past, see \cite{Jurdjevic1,Sachkov1} for a general overview, \cite{Bonnard2,Bonnard3,Jurdjevic2} for the particular case of affine fields.
Including the complete positivity requirements totally alter these results, because of the relaxation it induces.

The qualitative difference between studying the master equation and its controlled counterpart is that the master equation is an ODE whose solution, obtained integrating a single vector field, is a one parameter semigroup; the presence of control parameters in the controlled master equation implies that we have to consider a family of vector fields simultaneously, and therefore the admissible flow is a multiparameter semigroup or Lie semigroups \cite{Hilgert1}.
Such semigroup is the reachable set. 
When the reachable set is large enough to be a subgroup or at least to act transitively on the homogeneous space, then we have controllability.
The problems arise when the reachable semigroup is not a group, as in the case of the controlled master equation.
A novel element with respect to, for example, the control of Schr{\"{o}}dinger equations \cite{Cla-contr-root1} is that in the master equation one has to deal with a true {\em drift} term, i.e. a vector field which is both noncontrollable and nonrecurrent.
Then it can happen that although it is (often) possible to generate motion in any direction (i.e. we have the accessibility property), the system in never controllable in finite time because the flow cannot be reversed.
In other terms, the reachable set may be open and dense in the space of admissible density operators, but the initial condition of the controlled master equation always lies on the boundary of such set for any finite time and therefore it is not possible to reach arbitrary points in its neighborhoods.
The vector of coherences representation is very useful in this respect, as it allows to explain the lack of controllability in terms of the trace of the dissipation/relaxation superoperator.
In fact, the main reason for noncontrollability lies in the structure of the nonunitary operators given by complete positivity.
When such infinitesimal generators is unital this is clearly visible: for the density operator $ \rho $, $ \tr{\rho} $ gives the level sets of a quadratic Lyapunov function centered in the origin.
In this case, the controlled dynamics is stable and the control alone allows only to move within one of the level sets, not to pass from one level set to another.
Since the nonunitary operator is pointing inward, as time passes also the controlled integral curves can move only inward and this establishes a monotonicity relation among the sets reachable at different time instants.
As pointed out for example in \cite{Tannor1}, the presence of a dissipation (nonunitary) operator is essential for {\em any} motion not confined to a sphere in $ \mathbb{R}^n$ to be accomplished.
Notice that this holds regardless of the existence of a thermodynamic equilibrium, i.e. a fixed point for the original uncontrolled master equation.
For affine dissipation operators, the situation is slightly more complicated and controllability may be recovered as a limit process.
The atom with spontaneous emission is one such case and will be discussed in some detail.
In this case, motion is not confined to the inward of spheres in $ \mathbb{R}^n $ and ``purification'' processes are possible.

The organization of the paper is as follows: in Section~\ref{sec:drift-contr} we review all the relevant notions concerning controllability of bilinear/affine systems on homogeneous spaces of a Lie group; in Section~\ref{sec:ham-control} the formalism of the vector of coherences parameterization is recalled and used to discuss controllability of Liouville dynamics; in Section~\ref{sec:mas-control} the controllability of the master equation is treated and the main Theorem formulated.
Finally, in Section~\ref{sec:twol-control} the two-level case is discussed in detail, first for general dissipation operators and then for few significant examples.

It is worth remarking that all our considerations make sense for finite dimensional quantum systems.

\section{Drift and controllability for bilinear control systems}
\label{sec:drift-contr}
All properties introduced in this Section are standard in geometric control and are adequately surveyed for instance in \cite{Jurdjevic1,Sachkov1}.
Consider the following bilinear control system:
\beq
\begin{split}
\dot x & =  B_0 x + \sum_{k=1}^q u_k B_k x \\
x(0) & =  x_i
\end{split}
\label{eq:sys1}
\eeq
where the controls $ u_1, \ldots, u_q $ are real valued piecewise constant functions defined on $[0, \, \infty ) $, $ B_0, \ldots, B_q $ are square matrices and $ x \in M $, an analytic manifold of real dimension $n$. 
In this work: $M$ is $\mathbb{R}^n_0 = \mathbb{R}^n \setminus \{ 0 \} $ or some $n$-dimensional homogeneous space (like a sphere) contained in $ \mathbb{R}^r $, $r\geq n$, 
or some subset of $ \mathbb{R}^n $ like a solid unit ball.
The vector field $ B_0 x$ is called the drift and $ B_1 u_1 x , \ldots, B_q u_q x $ are the control vector fields.  

Given $ x_i \in M $, let us call $ {\cal R}(x_i, \, T ) $ the reachable set from $ x_i $ at time $ T> 0 $ for the system \eqref{eq:sys1} :
\[
{\cal R}  (x_i, \, T ) = \left\{ x \in M \: \text{s.t. $ x(0 ) = x_i $ and
   $  x(T) = x $, $T > 0 $, for some admissible control $ u_1, \ldots , u_q$} \right\}
\]
If $ {\cal R}  (x_i, \,\leq T ) = \cup_{ 0 \leq t \leq T} {\cal R} (x_i, \, t ) $, then the {\em reachable set} from $x_i $ is $ {\cal R}  (x_i) = \cup_{ 0 \leq t \leq \infty} {\cal R} (x_i, \, t ) $.
In correspondence of a given $ T_f $ and of ${\cal R}  (x_i, \, \leq T_f ) $, one can define the notions of finite time controllability as follows:
\begin{definition}
Given $T_f> 0$, the system \eqref{eq:sys1} is {\em $T_f$-controllable} if $ \forall \; x_i, \, x_f \in M $ $ \exists $ admissible control functions $ u_1 , \ldots u_q $ such that the flow of \eqref{eq:sys1} satisfies $ x(0) = x_i  $ and $x(T_f)  = x_f  $.
\end{definition}
The existence of a $T_f$ finite is important for the application discussed in this work. When, instead, we are interested in controllability for any time in ${\cal R}  (x_i) $, then we can use the following:
\begin{definition}
The system \eqref{eq:sys1} is {\em controllable} if any $ x_f \in M $ is reachable from any $ x_i \in M $ for some admissible control function $ u_1 , \ldots u_q $.
\end{definition}

Unlike the reachable set which accounts only for the positive time evolution of the trajectories of the system, the orbit $ {\cal O} (x_i ) $
requires to consider complete vector fields, i.e. defined on the whole
time axis:
\[ 
{\cal O} (x_i ) = \cup_{t \in \mathbb{R} }
  \left\{ x \in M \: \text{ s.t. $ x(0 ) = x_i $ and $ x(t) = x $, $ t\in \mathbb{R} $, for some admissible control $ u_1, \ldots , u_q$} \right\} 
\]

The difference between $ {\cal O}(x_i) $ and $  {\cal R}(x_i) $ corresponds to the difference between the accessibility and controllability properties.
\begin{definition}
The system \eqref{eq:sys1} is {\em accessible} if  $ {\cal R}  (x_i, \,\leq T ) $ contains nonempty open sets of $ M$ for all $ T> 0 $.
\end{definition}

While accessibility guarantees the existence of open reachable sets, it does not say anything on $ x_i $ belonging to it.
\begin{definition}
The system \eqref{eq:sys1} is {\em small-time controllable} if $ x_i $ belongs to the interior of the reachable set, $ {\rm int} \, {\cal R} (x_i, \,T ) $, for all $ T> 0$.
\end{definition}

The accessibility property admits an algebraic characterization in terms of the Lie algebra generated by the vector fields $B_0 x , B_1 x , \ldots, B_q x $, call it $ {\rm Lie}\left( B_0 x , B_1 x , \ldots B_q x  \right)$.
\begin{theorem}
\label{thm:larc}
(Lie algebraic rank condition (LARC))
The system \eqref{eq:sys1} is accessible if and only if $ {\rm dim }\left({\rm Lie}\left( B_0 x, B_1 x , \ldots B_q x \right) \right) = {\rm dim } \left( M \right) $.
\end{theorem}

For bilinear systems, when accessibility holds there exists a Lie group of transformations, call it $ G$, of (finite) dimension greater or equal than $ M$ acting transitively on $M$ and to which we can lift the system.
Invariance of the vector fields on a Lie group implies that the controllability conditions are global and independent of the point of application.
For example, for both $ {\cal R} (x_i) $ and $ {\cal O}(x_i) $ we have $ {\cal R} (x_i ) =   {\cal R}_G x_i $ and $ {\cal O} (x_i ) =   {\cal O}_G x_i $, with $ {\cal R}_G = {\cal R}(I) $ and $ {\cal O}_G = {\cal O}(I) $ reachable set and orbit of the lifted system, where $I$ is the identity matrix of $G$. Therefore we can work indifferently with vector fields on $ M$ ($ B_0 x , B_1 u_1 x, \ldots B_q u_q x $) or with right invariant vector fields on the Lie group $ G$ (i.e. the matrices $  B_0, B_1 u_1, \ldots B_q u_q $ of $\mathfrak{g}$, the Lie algebra of $G$), to which we have lifted the system, starting from the identity of $G$:
\beq
\begin{split}
\dot g  &  =  B_0 g + \sum_{i=1}^q  u_i (t) B_i  g
 \qquad g \in G    \\
 g(0 ) & =  I
\end{split} 
 \label{eq:sysLie2}
\eeq
In particular, the LARC condition and the so-called orbit theorem guarantee that $ {\cal O}_G$ is the whole $G $ and that $M$ is nothing but a homogeneous space of $G$ expressed in terms of equivalence classes as $ G x $, $ x \in M$. 
The Lie algebra $ \mathfrak{g} $ is therefore equal to $ {\rm Lie}\left(B_0, \ldots B_q \right) $ and the accessibility condition reformulates as transitivity of $G$ (or of $\mathfrak{g}$, with a common abuse of terminology) on $M$.
\begin{theorem}
\label{thm:acc-trans}
The system \eqref{eq:sys1} is accessible if and only if $ \mathfrak{g} $ is transitive on $ M $.
\end{theorem}
The LARC condition is only a necessary condition for controllability, even when it holds the reachable set needs not be the whole Lie group $G$.
When $ {\cal R}_G \subsetneq {\cal O}_G $ the lifted system is not controllable, the reason being that the drift is allowed to flow only along the time forward direction and may not be reversible by means of the control vector fields.
In fact, the control vector fields are ``complete'' in the sense that, since $ u_k $ can assume both positive and negative values, once exponentiated they generate a one parameter subgroup $  {\rm exp}\left( t  u_k  B_k , \; t \geq 0 \right) $.
On the contrary, the drift produces only a subsemigroup $  {\rm exp}\left( t B_0 , \; t\geq 0 \right) $ and thus $ {\cal R}_G $ in general only has the structure of a {\em Lie semigroup} of $G$ \cite{Hilgert1}.

The case of a compact group is exceptional, since compact Lie groups do not admit semigroups: $ {\rm exp}\left( t B , \; t\geq 0 \right) = {\rm exp}\left( t B , \; t \in \mathbb{R}  \right)$.  Hence $  {\cal R}_G $ collapses in $ {\cal O}_G$ and the accessibility property collapses into (``long time'') controllability.
The LARC condition then becomes necessary and sufficient for controllability: $ {\cal R} (x_i ) = {\cal O}_G x_i = G  x_i = M $, $ \forall \: x_i \in M $. 

In general, however, one has to deal with the case of $ {\cal R}_G $ being only a Lie semigroup. Even if $ {\cal R}_G$ is a proper semigroup, $ {\cal R}_G \subsetneq G $, it may still happen that the action of $ {\cal R}_G$ on $M$ is transitive. 
In the literature, most results are in the form of sufficient conditions for controllability. For the system \eqref{eq:sys1}, examples are 
\begin{enumerate}
\item $ {\cal R}_G = G $ and $G$ acts transitively on $M$;
\item $ {\cal R}_G $ acts transitively on $M$;
\item $ x_i \in {\rm int}\, {\cal R}(x_i, \, T ) $ $ \forall \, T>0$
\end{enumerate}
For our case, none of these (or similar) conditions hold and ``negative'' results have to be established.

\subsection{Affine vector fields case}
The case of affine vector fields generalizes \eqref{eq:sys1} to the following set of ODEs:
\beq
\begin{split}
\dot x & =  B_0 x +  b_0 x_0 + \sum_{k=1}^q \left( u_k B_k x + b_k x_0 \right)  \\
x(0) & =  x_i
\end{split}
\label{eq:sys-aff1}
\eeq
where $x_0 $ is a real constant.
It corresponds to a Lie group of transformations having the structure of a semidirect product $K \circledS V $ with $V$ typically a $n$-dimensional real vector space and $K$ a Lie group acting linearly on it. 
The dimension of $ K \circledS V$ is $ {\rm dim} \left( K \right) + n $.
By choosing the following homogeneous coordinates for the state $ \bar{x} = [x_0 , \,  x^T ]^T$, the system \eqref{eq:sys-aff1} recovers the linear form of \eqref{eq:sys1}:
\[
\dot{\bar{x}}  =  \bar{B}_0 \bar{x} + \sum_{k=1}^q u_k \bar{B}_k \bar{x} \\
\label{eq:sys-aff2}
\]
where $ \bar{B}_k=\begin{bmatrix} 0 & 0 \\ b_k & B_k \end{bmatrix} $.
The homogeneous coordinates allow to transform the affine action of $ K\circledS V $ on $ x $ into linear action on $ \bar{x}$.
If $ g = \begin{bmatrix} 1 & 0 \\ v & k \end{bmatrix} \in G = \begin{bmatrix} 1 & 0 \\ V & K \end{bmatrix} $, the action $ \Phi \, : \, G \times M \to M $ is:
\[
\Phi(g) (\bar{x} ) = g \bar{x} = \begin{bmatrix} x_0 \\ k x + v x_0 \end{bmatrix}
\]
so that the affine vector fields induced on $\bar{x} $ by $ \Phi$ are:
\[
\Phi_\ast (\bar{B} ) (\bar{x} ) = \bar{B} \bar{x} = \begin{bmatrix} 0 \\ Bx + b x_0 \end{bmatrix}
\]
and the Lie bracket is:
\[
\left[ \bar{A}, \, \bar{B} \right] =\begin{bmatrix} 0 & 0 \\ Ab-Ba & [A, \, B] \end{bmatrix} 
\]
Also for affine systems, special sufficient conditions for accessibility and controllability  have been devised, see \cite{Jurdjevic2,Bonnard2,Bonnard3} for details.

\section{Controllability of Hamiltonian dynamics}
\label{sec:ham-control}
To describe differential equations for density operators, we make use of the so-called vector of coherences formulation.
A few essential facts about it are reported below; see for example \cite{Alicki1} for a thorough description and further references.

\subsection{Density operators and vectors of coherences}
The state of a quantum mechanical system in an $N$-dimensional Hilbert space $ {\cal H}^N $ can be described in terms of a trace 1 positive semidefinite Hermitian operator $ \rho$ called the density matrix.
If the density operator is entirely characterized by a wavefunction $ \ket{\psi }$, then the system is said to be in a pure state, $\rho$ is defined as $ \rho(t) = \ket{\psi} \bra{\psi} $ and $ {\rm tr } (\rho^2 ) = 1 $. If instead we have a statistical ensemble $ \rho(t) = \sum_{i=1}^N p^{(i)} \ket{\psi^{(i)}} \bra{\psi^{(i)}} $ for $  p^{(i)} \geq 0$ and $ \sum_{i=1} ^N  p^{(i)} =1 $, then the system is said to be in a mixed state characterized by the pairs $\{ p^{(i)}, \, \ket{\psi^{(i)}} \}$ and $ {\rm  tr} (\rho^2 ) \leq 1 $.
In both cases, the properties of Hermitianity $ \rho = \rho^\dagger $ and of unit trace $ {\rm tr} (\rho )  =1 $ imply that the $ N \times N $ matrix representing the density operator depends on $n = N^2 -1 $ real parameters.
Up to the imaginary unit, $N\times N  $ traceless Hermitian matrices form the Lie algebra $ \mathfrak{su}(N) $ of dimension exactly $n $.
If to it we add the (properly normalized) unit vector, then we obtain a complete basis for the density operator of an $N$-dimensional quantum mechanical system.
In fact, the $N$-dimensional Pauli matrices $ \lambda_j $, see for example \cite{Lichtenberg1} for their explicit expression, and the identity matrix $ \lambda_0 = N^{-\frac{1}{2}} I $, form a complete orthonormal set of basis operators for $ \rho $ (orthonormal in the sense that $ {\rm tr} (\lambda_j \lambda_k ) = \delta_{jk} $).
In particular, then, $ \rho = \sum_{j=0} ^{n } {\rm tr }( \rho \lambda_j ) \lambda_j = \sum_{j=0} ^{n }  \rho_j \lambda_j $, with $ \rho_0 = N^{-\frac{1}{2} } $ fixed constant and the $n $ real parameters $ \rho_j $ giving the parameterization of $\rho$.
Since the $ \lambda_k $, $ k=1, \ldots, n$, form a compete set of observable operators, the $ \rho_j = {\rm tr}( \rho \lambda_j )  $ are expectation values of $ \rho$.
Call $ \bm{\rho}= [ \rho_1 \ldots \rho_n ]^T $ such {\em vector of coherences} of $ \rho $.
Due to the constant component along $ \lambda_0 $, $ \bm{\rho} $ is living on an affine space characterized by the extra fixed coordinate $ \rho_0 = N^{-\frac{1}{2} } $. Such $n$ dimensional Liouville space of vectors $ \bar{\bm{\rho}} =  [ \rho_0 \, \rho_1 \ldots \rho_n ]^T = [ \rho_0 \, \bm{\rho}^T ]^T $ has Euclidean inner product given by the trace metric: $ \|  \bar{\bm{\rho}} \| =\sqrt{ \llangle \bar{\bm{\rho}}, \, \bar{\bm{\rho}} \rrangle}=  \sqrt{\tr{\rho^2}} $.
The condition $  {\rm  tr} (\rho^2 ) \leq 1 $ then translates in $  \bar{\bm{\rho}} $-space as $  \bar{\bm{\rho}} $ belonging to the solid ball of radius $ 1 $ centered at $  [ \rho_0 \, 0 \, \ldots\, 0 ]^T $, call it $ \bar{\mathbb{B}}^n $, for all positive times.
The surface of such ball generalizes the idea of Block sphere to $N$ dimension and corresponds to pure states $ \| \bar{\bm{\rho}} \|^2 = 1 $.
In terms of vector of coherences, the condition $ \tr{\rho^2 }\leq 1 $ becomes the smaller ball $ \| \bm{\rho} \| ^2 \leq 1- \frac{1}{N} $ (centered at the origin).

\subsection{Hamiltonian dynamics}

If $ H$ is a constant finite dimensional Hamiltonian, for the density operator the Liouville equation is given by 
\[
\dot \rho (t) = -i [ H , \, \rho ] = - i {\rm  ad }_H (\rho) 
\label{eq:dens1}
\]
If $ -i H \in \mathfrak{su}(N) $, then $ -i {\rm ad}_H $ is a so-called commutator superoperator i.e. a linear operator in the $n $ dimensional Liouville space of $ \bm{\rho} $ vectors. 
In terms of $ \bm{\rho}$, the action of $ -i {\rm ad}_H $ is linear:
\beq
\dot{\bm{\rho}} =- i  {\rm ad}_H \, \bm{\rho}
\label{eq:dens-lin}
\eeq
$ H $ being traceless and Hermitian, in the $ \left\{ \lambda_j  \right\} $ basis: $ H = \sum_{l=1}^n h_l \lambda_l $. 
The process of passing from $ \rho $ to $ \bm{\rho} $ is mathematically equivalent to passing to the adjoint representation of the Lie algebra $ \mathfrak{su}(N) $.
In fact, the corresponding basis in the adjoint representation is given by the $ n \times n $ matrices $ {\rm ad}_{\lambda_1} , \ldots   {\rm ad}_{\lambda_n} $ of elements $ \left( {\rm ad}_{\lambda_l} \right)_{jk} =i f_{ljk} $ with $ f_{ljk} $ real fully skew-symmetric (with respect to the permutation of any pair of indexes) tensor.
Thus $ - i {\rm ad}_H = -i \sum_{l=1}^n h_l {\rm ad}_{\lambda_l} $.
The $n\times n $ matrices $ -i {\rm ad}_{\lambda_1}, \ldots ,  -i {\rm ad}_{\lambda_n} $ are real and skew-symmetric and as such they are part of a basis of $ \mathfrak{so}(n) $.
Since $ \dim \left(\mathfrak{so}(n) \right) = \frac{n(n-1)}{2} = \frac{N^4 - 3 N^2 + 2 }{2} $, for $N> 2 $ the $n$ matrices $ -i {\rm ad}_{\lambda_1}, \ldots ,  -i {\rm ad}_{\lambda_n} $ span only a proper subalgebra $ {\rm ad}_{\mathfrak{su}(N) } $ of $\mathfrak{so}(n)$.
For example for $ N=3 $, $ n = \dim \left( \mathfrak{su}(3) \right) =8 $, while $ \dim \left( \mathfrak{so}(8) \right) = 28 $!
In the Liouville equation \eqref{eq:dens-lin}, the propagator for $ \bm \rho$ corresponding to the Hamiltonian $H$ is an orthogonal matrix:
\[
\bm \rho (t)  = g(t) \bm \rho(0), \qquad  g(t) \in{\rm exp}({\rm ad}_{\mathfrak{su}(N)} ) \subseteq SO(n)
\]
such that $ \dot g(t) = -i {\rm ad}_H g(t)$, $ g(0) = I $. 
The action of any $ g\in SO(n) $ on $ \rho$ is isometric and as such it preserves the inner product $ \| \bm{\rho} \| $.

\subsection{Coherent control of Hamiltonian dynamics}
\label{sec:trans-unit}
Assume that the Hamiltonian $H$ is composed of a time-invariant part $H_0$ representing the free evolution of the system plus $q$ time-varying forcing terms representing the interaction with $q$ external fields, modeled semiclassically:
\[
 H(t) =  H_0 + \sum_{k=1}^q u_k(t) H_k, \qquad -i H_0, -i H_k \in \mathfrak{su}(N) 
\]
with the parameters $ u_k $ representing the control fields.
Consider a pure state of ket $ \ket{\psi} \in \mathbb{S} \subset {\cal H} ^N $ (where $ \mathbb{S} $ is the sphere in $ N$ dimensional Hilbert space) and its Schr{\"{o}}dinger equation 
\beq
i \dot{\ket{\psi}} = H_0 \ket{\psi} + \sum_{k=1}^q u_k H_k \ket{\psi} \quad \ket{\psi(0) } = \ket{\psi_0 }
\label{eq:schrod1}
\eeq
The sphere $ \mathbb{S} $ in $ {\cal H}^N $ is a homogeneous space of $ SU(N)$. Compactness of $ SU(N)$ plus transitivity of the $SU(N)$ action on $ \mathbb{S} $ in this case guarantee the following (see \cite{Jurdjevic3} for the original formulation or \cite{Cla-contr-root1} for a thorough discussion):
\begin{theorem} 
\label{thm:jurd1}
The system \eqref{eq:schrod1} is controllable if and only if ${\rm Lie}\left( -i H_0, \, -i H_1, \ldots , -i H_q \right) = \mathfrak{su}(N) $ \end{theorem}
By computing the (real) dimension of such Lie algebra: $ {\rm dim}\left(  {\rm Lie}\left( -i H_0,  \ldots , -i H_q \right) \right)$ $ = {\rm dim}\left( \mathfrak{su}(N)\right) $ $  = n = {\rm dim}\left( \mathbb{S} \right) $.

In the following we will always consider the controllable case for the wavefunction $ \ket{\psi}$:

$ \; $ \\ \noindent
{\bf{Assumption A1}}
{\em The system \eqref{eq:schrod1} is controllable.}

$ \; $ \\ \noindent
Passing to density matrices, for a mixed state $ \rho $ driven by the same Hamiltonian $ H(t)$ the corresponding forced Liouville equation written in terms of vector of coherences is:
\beq
\dot{\bm{\rho}} = -i {\rm ad}_{H_0} \bm{\rho} -i \sum_{k=1}^q u_k {\rm ad}_{H_k} \bm{\rho}
\label{eq:liouv1}
\eeq
The vector fields $ -i {\rm ad}_{H_0} , \ldots -i {\rm ad}_{H_q}  $ corresponding to the Hamiltonian dynamics lack the translation component and belong to a subalgebra  ${\rm ad}_{\mathfrak{su}(N)} $ of $\mathfrak{so}(n) $.
Just like the Lie group $SU(N)$ is acting transitively on the unit sphere on $ {\cal H}^N $, so the orthogonal group $ SO(n)$ is acting transitively on a sphere $ \| \bm{\rho} \|^2 = {\rm const} \leq 1 - \frac{1}{N} $.
By dimension counting, $ {\rm exp}({\rm ad}_{\mathfrak{su}(N)} ) $ is not acting transitively on such sphere if $N > 2 $.
In fact, it is well-known that coherent control cannot modify the eigenvalues of $ \rho$, and so controllability can occur only inside the leaf of the foliation of $ \rho $ (determined by the eigenvalues) that one starts with. See \cite{Schirmer6} for a description of the kinematic equivalence classes of density matrices in the context of dynamical control, or \cite{Lendi1} for a complete description of the invariants of motion.

\section{Controllability of Markovian master equations}
\label{sec:mas-control}
The requirement of $ \tr{\rho^2} \leq 1 $ for the density operator is reformulated in the vector of coherences parameterization as $ \| \bar{\bm{\rho}}\|^2 \leq 1 $. Thus $ \bar{\mathbb{B}}^n $ has to be made invariant by the quantum dynamical evolution.
The main feature of the master equation is to capture all the possible infinitesimal generators that fulfill this condition.
Obviously, also the driven master equation has to live on $ \bar{\mathbb{B}}^n$, and all controllability questions have to be restricted to $ \bar{\mathbb{B}}^n$.

\subsection{Master equation}

Calling $ {\cal L}_D $ the superoperator representing the relaxing/dissipating part of the dynamics, in the basis $\{ \lambda_j \}$ of traceless Hermitian matrices the Markovian master equation is expressed as \cite{Gorini1}:
\beq
\begin{split}
\dot \rho & ={ \cal L}_H (\rho) + {\cal L}_D (\rho) = -i {\rm ad}_H (\rho) + \frac{1}{2} \sum_{j, k =1 }^n a_{jk} \left( [\lambda_j , \, \rho \lambda_k ] + [ \lambda_j \rho , \, \lambda_k ] \right) \\
& =  -i [ H, \, \rho ] +\frac{1}{2} \sum_{j, k =1 }^n a_{jk} \left( 2 \lambda_j \rho \lambda_k - \{ \lambda_k \lambda_j , \, \rho \} \right)
\end{split}
\label{eq:master1}
\eeq
where the Hermitian matrix $ A = ( a_{jk} )$ is positive semidefinite, $A  \geq 0 $, and $ \{ \, \cdot \, , \, \, \cdot \, \} $ is the anticommutator.
For the basis $\{ \lambda_j \}$, unlike a Lie bracket which is linear in the generators, the anticommutator has an affine structure: $
\{ \lambda_j , \, \lambda_k \} = \frac{2\sqrt{2} }{N} \delta_{jk} \lambda_0 + \sum_{l=1}^n d_{jkl} \lambda_l$,
with $ d_{jkl} $ the real and fully symmetric tensor.
The expression of \eqref{eq:master1} in terms of vector of coherences is as follows:
\beq
\dot{\bm{\rho}} = -i {\rm ad}_H \bm{\rho} + \sum_{j,k=1}^n a_{jk} \left( L_{jk} \bm{\rho} + \bm{v}_{jk} \rho_0 \right)
\label{eq:masterlin1}
\eeq
with $ L_{jk}$ $ n\times n$ complex matrix of mixed symmetry and $ \bm{v}_{jk} $ imaginary $n$ vector given by:
\beqa
L_{jk} & = & \left( L_{jk} \right)_{lr}  =  -\frac{1}{4} \sum_{m=1}^n  \left( ( f_{jmr} + i d_{jmr} ) f_{kml} + ( f_{kmr} - i d_{kmr} ) f_{jml} \right) 
\label{eq:Ljk} \\
\bm{v}_{jk} & = & \frac{i}{\sqrt{N}}  [ f_{jk1} \ldots f_{jkn} ] ^T
\nonumber
\eeqa

\subsection{Coherent control of master equations}

Under the assumption of weak and high frequency control fields, it is acceptable to assume that no time dependence is induced in the $ {\cal L}_D $ term by the external fields.
Adding the controls, eq. \eqref{eq:masterlin1} modifies as \footnote{In general, $ H_0 $ may not be the same as in \eqref{eq:liouv1}, as is well-known. This is irrelevant for our purposes.}:
\beq
\begin{split}
\dot{\bm{\rho}}  & = {\cal L}_{H_0} \bm{\rho} + \sum_{k=1}^q u_k {\cal L}_{H_k} \bm{\rho} +  {\cal L}_D \bm{\rho} \\
& = -i {\rm ad}_{H_0} \bm{\rho} -i \sum_{k=1}^q u_k {\rm ad}_{H_k} \bm{\rho} + \sum_{j,k=1}^n a_{jk} \left( L_{jk} \bm{\rho} + \bm{v}_{jk} \rho_0 \right)
\end{split}
\label{eq:masterlin2}
\eeq
or, in homogeneous coordinates and calling $ \bar{L}_{jk} = \begin{bmatrix} 0 & 0 \\ \bm{v}_{jk} & L_{jk} \end{bmatrix}$, $ j, k = 1, \ldots, n $:
\beq
\begin{split}
\dot{\bar{\bm{\rho}}}  & = \bar{{\cal L}} \bar{\bm{\rho}} = \bar{{\cal L}}_{H_0} \bar{\bm{\rho}} + \sum_{k=1}^q u_k \bar{{\cal L}}_{H_k} \bar{\bm{\rho}} +  \bar{{\cal L}}_D \bar{\bm{\rho}} \qquad \qquad \bar{\bm{\rho}} \in \bar{\mathbb{B}}^n\\
& = \begin{bmatrix} 0 & 0 \\ 0 & -i {\rm ad}_{H_0} \end{bmatrix} \bar{\bm{\rho}} 
+ \sum_{k=1}^q u_k \begin{bmatrix} 0 & 0 \\ 0 & -i {\rm ad}_{H_k} \end{bmatrix} \bar{\bm{\rho}} 
+ \sum_{j,k = 1 }^n a_{jk} \bar{L}_{jk} \bar{\bm{\rho}} 
\end{split}
\label{eq:masterlin3}
\eeq
The state of \eqref{eq:masterlin3} is living on $ \mathbb{R}^{n+1}$ and is constrained by the positivity of $ \rho$ requirement to belong to $ \bar{\mathbb{B}}^n$.
However, the dissipation term $ {\cal L}_D $ is not coherent and as such it enlarges the integral group of \eqref{eq:masterlin3} from $ {\rm exp}\left( {\rm ad}_{\mathfrak{su}(N)} \right) $ to one of the Lie groups properly containing it.
Examples are $ SO(n,\, \mathbb{R}) $, $ SL(n,\, \mathbb{R}) $, $ GL^+ (n, \, \mathbb{R} ) $, the connected component of $ GL(n, \, \mathbb{R} ) $ containing the identity, or their semidirect extensions $ SO(n,\, \mathbb{R} )\circledS \mathbb{R}^n  $, $ SL(n,\, \mathbb{R} )\circledS \mathbb{R}^n  $ and $ GL^+ (n, \, \mathbb{R} )\circledS \mathbb{R}^n  $.

Since $A$ is Hermitian, the number of independent parameters $ a_{jk}$ is $ n^2 $.
It is convenient to rearrange the $ n^2 $ degrees of freedom in the following manner.
From \eqref{eq:Ljk}, it is straightforward to check that $ {\rm Re} \left[ ( L_{kj} ) _{lr} \right] = {\rm Re} \left[ ( L_{jk} ) _{lr} \right] $, $ {\rm Im} \left[ ( L_{kj} ) _{lr} \right] =- {\rm Im} \left[ ( L_{jk} ) _{lr} \right] $ and therefore $ L_{kj} = L_{jk}^\ast $.
If we call $ L_{jk}^\Re =  {\rm Re} \left[ ( L_{jk} ) _{lr} \right] $ and  $ L_{jk}^\Im =  {\rm Im} \left[ ( L_{jk} ) _{lr} \right] $, then with respect to index permutation $  L_{jk}^\Re $ is symmetric, $  L_{kj}^\Re =  L_{jk}^\Re $ while $  L_{jk}^\Im $ is skew-symmetric, $  L_{kj}^\Im = - L_{jk}^\Im $.
Similarly, the Hermitianity of $ A$ implies $ a_{kj} = a_{jk}^\ast$ or, if we write $  a^\Re_{jk}= {\rm Re} [a_{jk}] $ and $ a_{jk}^\Im = {\rm Im} [a_{jk}] $,
$ a_{kj}^\Re = a_{jk}^\Re $, $ a_{kj}^\Im  = - a_{jk}^\Im $. 
Therefore:
\beq
a_{jk} \bar{L}_{jk} + a_{kj} \bar{L}_{kj}  = ( 2 - \delta_{jk} ) a_{jk}^\Re \begin{bmatrix}  
0 & 0 \\ 
0 &   L_{jk}^\Re 
\end{bmatrix}
+ 2 a_{jk}^\Im 
\begin{bmatrix}  0 & 0 \\
i \bm{v}_{jk} & - L_{jk}^\Im 
\end{bmatrix}, \qquad 1 \leq j \leq k \leq n 
\label{eq:ajkakj}
\eeq
To have $ A\geq 0 $, a number of constraints among the $ a_{jk} $ must be imposed like, for example, $ a_{jj} = a_{jj}^\Re \geq 0 $ and $ a_{jj} a_{kk} \geq ( a_{jk}^\Re ) ^2 + (a_{jk}^\Im )^2 $ or $ | a_{jk}| \leq ( a_{jj} + a_{kk} )/2 $.

Our aim here is to draw conclusions about which $ \bar{\bm{\rho}}$ can be reached by means of coherent control.
In \eqref{eq:masterlin3}, unlike $  \sum_{k=1}^q u_k \bar{{\cal L}}_{H_k}$, both $  \bar{{\cal L}}_{H_0} $ and $   \bar{{\cal L}}_D $ have integral curves that can flow only along the positive semiorbit and, in control terms, $ \bar{{\cal L}}_{H_0} $ plays the role of the drift and $ \bar{{\cal L}}_D $ that of a disturbance.
Classically, a disturbance can be treated for example as a parametric family of vector fields with parameters belonging to admissible intervals (here $ a_{jk} $ such that $A \geq 0$). 
However, in the case of $ \bar{{\cal L}}_D $ parametric the master equation becomes a differential inclusion and little can be said about its controllability properties.
Therefore in this work we will assume to be dealing only with a precisely known value of $A$ of hence of $ \bar{{\cal L}}_D $.

$ \; $ \\ \noindent
{\bf Assumption A2}
{\em The parameters $ a_{jk} $, $ j, k=1, \ldots, n $ are fixed and known exactly.} 

$ \; $ \\ \noindent 
Consequently we can treat $ \bar{\cal L}_D $ as a part of the drift term (together with $  \bar{{\cal L}}_{H_0} $) and use the tools of Section~\ref{sec:drift-contr}.

Under the assumption of unitary controllability, the Lie algebra $ \mathfrak{g} $ of interest here is the smallest Lie algebra of real matrices containing $ {\rm ad}_{\mathfrak{su}(N) } $ and $  \bar{{\cal L}}_{H_0} +\bar{\cal L}_D $ and closed with respect to matrix commutation:
\[
\mathfrak{g} = {\rm Lie} \left(  \bar{{\cal L}}_{H_0} +\bar{\cal L}_D , \,  \bar{{\cal L}}_{H_1}, \ldots ,  \bar{{\cal L}}_{H_q} \right)
\]

Once $\mathfrak{g} $ is computed, the system can be lifted to $G$:
\beq
\dot{g} = \bar{{\cal L}}_{H_0}g + \sum_{k=1}^q u_k \bar{{\cal L}}_{H_k} g +  \bar{{\cal L}}_D g
\label{eq:masterlinlift}
\eeq

The following Theorem gathers various results about accessibility and controllability for system \eqref{eq:masterlin3}.
Concerning controllability, while for $ \bar{\cal L}_D $ unital the results are sharp (and negative), the case of $ \bar{\cal L}_D $ affine is more difficult to treat.
In fact, in this case, in spite of the lack of small-time controllability it may happen that points that are not reachable in short time are reachable for $T$ large enough and even that $ {\rm cl}\left({\cal R}(x_i ) \right) = \bar{\mathbb{B}}^n $ asymptotically (${\rm cl}(\, \cdot \, ) $ means closure).
The atom with spontaneous emission discussed in the examples of Section~\ref{subs:examples} is one such case.
Essentially this fact depends on the existence of a fixed point for the master equation and on it being on the boundary of $ \bar{\mathbb{B}}^n$, $ \partial \, \bar{\mathbb{B}}^n$.
However, even in this case $ \partial \, \bar{\mathbb{B}}^n$ is reached only asymptotically and therefore the system fails to be controllable in finite time.

\begin{theorem}
\label{thm:acc-contr-N}
Under Assumptions {\bf A1} and {\bf A2}, we have the following:
\begin{enumerate}
\item the system \eqref{eq:masterlin3} is accessible in $ \bar{\mathbb{B}}^n $ if and only if $ \mathfrak{g} = \mathfrak{gl} ( n , \mathbb{R} )$ or $ \mathfrak{g} = \mathfrak{gl} ( n , \mathbb{R} ) \circledS \mathbb{R}^n $ 
\item the system \eqref{eq:masterlin3} is never small-time controllable in $ \bar{\mathbb{B}}^n $ for $ \bar{\cal L}_D \neq 0 $.
\item the system \eqref{eq:masterlin3} is never $ T_f$-controllable in $ \bar{\mathbb{B}}^n $ for any $ T_f > 0 $ and $ \bar{\cal L}_D \neq 0 $.
\item the system \eqref{eq:masterlin3} is never controllable in $ \bar{\mathbb{B}}^n $ for $ \bar{\cal L}_D \neq 0 $ unital.
\end{enumerate}
\end{theorem}

\proof

The proof of Part 1 follows from Theorem~\ref{thm:acc-trans}. The only transitive Lie algebras on $ \mathbb{R}^n_0 $ (and thus on $\bar{\mathbb{B}}^n$) are $ \mathfrak{sl}(n, \, \mathbb{R} ) $ and $ \mathfrak{gl}(n, \, \mathbb{R} ) $ and their semidirect extensions $ \mathfrak{sl}(n, \, \mathbb{R} )\circledS \mathbb{R}^n $ and $ \mathfrak{gl}(n, \, \mathbb{R} )\circledS \mathbb{R}^n $.
Recall that matrices in $  \mathfrak{sl}(n, \, \mathbb{R} ) $ are traceless and that, using the decomposition $\mathfrak{gl}(n, \, \mathbb{R} ) = \mathfrak{sl}(n, \, \mathbb{R} ) \oplus {\rm span} (I_n) $ ($ I_n $ the $ n \times n $ identity matrix), if $ {\rm tr}(\bar{\cal L}_D) = n \alpha$ and $ \bar{I} = \begin{bmatrix} 0 & 0 \\ 0 & I_n \end{bmatrix}$, $ \bar{\cal L}_D $ can be split as $ \bar{\cal L}_D = \alpha \bar{I} +  \tilde{\cal L}_D $, $\alpha\in \mathbb{R} , \, \alpha< 0 $, $ \tilde{\cal L}_D \in  \mathfrak{sl}(n, \, \mathbb{R} )\circledS \mathbb{R}^n $.
But since $ a_{jj}\geq 0 $, in order for $ \bar{\cal L}_D $ to be traceless it must be $ a_{jj} = 0 $ $ \forall \, j =1, \ldots , n$ and hence, from $ | a_{jk} | \leq ( a_{jj} + a_{kk} ) /2$ all $ a_{jk} = 0 $.
Therefore only $ \mathfrak{gl}(n, \, \mathbb{R} ) $ and $ \mathfrak{gl}(n, \, \mathbb{R} )\circledS \mathbb{R}^n $ are compatible with $ A \geq 0 $.

To prove Part 2, one needs to show that the initial condition $ \bar{\bm{\rho}}_i $ does not lie in $ {\rm int}\left( {\cal R}( \bar{\bm{\rho}}_i) \right) $. 
It is quite easy to verify it for $  \bar{\cal L}_D $ unital.
In fact, if the initial state $ \bar{\bm{\rho}}_i  $ is such that $ 0 < \| \bar{\bm{\rho}}_i \| = \delta \leq 1 $, then for the Hamiltonian part $ \llangle \left( \bar{{\cal L}}_{H_0} + \sum_{k=1}^q u_k \bar{{\cal L}}_{H_k} \right)  \bar{\bm{\rho}}_i , \,  \bar{\bm{\rho}}_i \rrangle =0 $ while $  \bar{\cal L}_D $ is pointing inward: $ \llangle  \bar{\cal L}_D  \bar{\bm{\rho}}_i , \,  \bar{\bm{\rho}}_i \rrangle < 0 $.
Therefore, the ball of radius $ \delta $ is invariant for the flow of \eqref{eq:masterlin3} and $  \bar{\bm{\rho}}_i  $ lies on the boundary of $ {\cal R}( \bar{\bm{\rho}}_i)  $.
For $ \bar{\cal L}_D $ affine, the lack of small-time local controllability is automatically verified for pure states $ \| \bar{\bm{\rho}}_i \|=1 $, because the physics of the problem imposes that $  \bar{\bm{\rho}} $ such that  $ \| \bar{\bm{\rho}} \|= \sqrt{1+ \epsilon }$, $ \epsilon > 0 $ is not admissible.
Writing the integral curves of the control system as $  \bar{\bm{\rho}} (t) = \Phi\left( {\cal T }{\rm exp} \int_0^t (\bar{\cal L} (\tau) d \tau \right) ( \bar{\bm{\rho}}_i) = g(t)  \bar{\bm{\rho}}_i  $ with ${\cal T}$ the Dyson operator, we can lift the dynamics to the system \eqref{eq:masterlinlift} with initial condition $ g(0) =I$. 
$   \bar{\bm{\rho}}_i \notin {\rm int} \left( {\cal R}( \bar{\bm{\rho}}_i ) \right) $ for $ \| \bar{\bm{\rho}}_i \|=1 $ implies that the reachable set $  {\cal R}( \bar{\bm{\rho}}_i ) =  {\cal R}_G \bar{\bm{\rho}}_i  $ cannot be transitive on any neighborhood of $ \bar{\bm{\rho}}_i $ and that for the lifted system $ I \notin {\rm int} \left( {\cal R}_G \right)  $.
But, due to right invariance, the properties of accessibility, controllability and transitivity have a global character and therefore $ {\cal R}_G $ is not transitive for any neighborhood of any $ \bar{\bm{\rho}}_i \in \bar{\mathbb{B}}^n $.

Concerning Part 3, if a finite time $ T_f $ is fixed, the reachable set $ {\cal R}(\bar{\bm{\rho}}_i, \, \leq T_f ) $ for the master equation is always only a Lie semigroup.
In fact, if the fixed point of $ \bar{\cal L}_D $ (when it exists) belongs to $ {\rm int} \left( {\cal R}(\bar{\bm{\rho}}_i, \, \leq T_f ) \right) $ then $ {\rm cl} \left({\cal R}(\bar{\bm{\rho}}_i , \, \leq T_f )\right) \subseteq {\rm cl} \left( {\cal R}(\bar{\bm{\rho}}_i ) \right) \subsetneq \bar{\mathbb{B}}^n $; if instead it belongs to $ \partial \,\bar{\mathbb{B}}^n$ then $ {\rm cl} \left({\cal R}(\bar{\bm{\rho}}_i,  \, \leq T_f )\right) \subsetneq  {\rm cl} \left( {\cal R}(\bar{\bm{\rho}}_i )  \right) = \bar{\mathbb{B}}^n $.
Even if a fixed point does not exist, we have that the norm of $ \bar{\bm{\rho}}_i $ can grow only if $ \llangle  \bar{\cal L}_D  \bar{\bm{\rho}}_i , \,  \bar{\bm{\rho}}_i \rrangle > 0 $ and that $ \| \bar{\bm{\rho}}(t) \| $ can approach $1$ at most as $ t \to \infty $.
Since $ \bar{{\cal L}}_{H_0}$ and $  \bar{{\cal L}}_{H_1}, \ldots   \bar{{\cal L}}_{H_q} $ preserve the length, excluding the trivial cases the control cannot speed up the convergence to $ \partial \, \bar{\mathbb{B}}^n$ from its ``best'' initial condition. But even in that case convergence is only asymptotic. 
Therefore, for any fixed $ T_f $ the open set $ {\cal R}( \bar{\bm{\rho}}_i , \, \leq T_f )$ cannot be all of $ \bar{\mathbb{B}}^n$ and neither can its closure.

Finally, the proof of noncontrollability in $ \bar{\mathbb{B}}^n$ for $ \bar{\cal L}_D $ unital follows from the same argument used above in Part 2.

\qed

For the system lifted to its integral group, the small-time controllability property collapses into controllability and we have the following.

\begin{corollary}
The ``lifted'' system \eqref{eq:masterlinlift} is accessible for $G = GL^+ (n, \, \mathbb{R} ) $ or $G = GL^+ (n, \, \mathbb{R} )\circledS \mathbb{R}^n $ but it is never controllable on $G$ for $ \bar{\cal L}_D \neq 0 $.
\end{corollary}

\proof
The first part is obvious, since accessibility on the Lie group is a necessary condition for accessibility on the homogeneous space.
Concerning controllability, from the proof of Theorem~\ref{thm:acc-contr-N}, Part 2, for the system \eqref{eq:masterlinlift}, $ I \notin {\rm int}\left( {\cal R}_G \right) $.
But, for Lie groups, such property is a global one and therefore the systems in never controllable.
\qed

Another way to prove the previous Corollary is via piecewise constant controls: in this case, $ g(t)= {\cal T }{\rm exp} \int_0^t \bar{\cal L} (\tau) d \tau =\prod_{j=1}^r {\rm exp}\left( \left( \bar{{\cal L}}_{H_0} + \bar{\cal L}_D + \sum_{k=1}^q u_{k_j} \bar{{\cal L}}_{H_k} \right) ( t_j - t_{j-1} ) \right) $ and using the formula $ \det ( {\rm exp} ( \cdot ) ) = {\rm exp}( \tr{\cdot } ) $ we have 
$ \det \left( \prod_{j=1}^r {\rm exp}\left( \left( \bar{{\cal L}}_{H_0} + \bar{\cal L}_D + \sum_{k=1}^q u_{k_j} \bar{{\cal L}}_{H_k} \right) ( t_j - t_{j-1} ) \right) \right) = {\rm exp} \left( \sum_{j=1}^r {\rm tr} \left( \left( \bar{{\cal L}}_{H_0} + \bar{\cal L}_D + \sum_{k=1}^q u_{k_j} \bar{{\cal L}}_{H_k} \right) ( t_j - t_{j-1} ) \right) \right) =  {\rm exp}\left( \tr{ \bar{\cal L}_D } t \right) \leq 1 $.
Therefore, $ g(t) \in  GL^+ (n, \, \mathbb{R} ) $ or $g(t) \in GL^+ (n, \, \mathbb{R} )\circledS \mathbb{R}^n $ (only in these cases \eqref{eq:masterlinlift} is accessible) must be such that $ \det ( g(t)) \leq 1 $ and cannot generate the whole Lie group.

Yet another method to show the same thing is to use the necessary condition of Lemma 6.8 of \cite{Jurdjevic3}. Write $ \bar{\cal L}_D = \alpha \bar{I} +  \tilde{\cal L}_D $, $\alpha\in \mathbb{R} , \, \alpha< 0 $, $ \tilde{\cal L}_D \in  \mathfrak{sl}(n, \, \mathbb{R} )\circledS \mathbb{R}^n $.
Since $ [ \bar{I}, \, \bar{F}]=0 \; \forall \, \bar{F} \in \mathfrak{gl}(n, \, \mathbb{R} )\circledS \mathbb{R}^n$, under the assumption of accessibility the ideal in $ \mathfrak{g} $ generated by the control vector fields coincides with the derived subalgebra $ \mathfrak{g}_1 = [ \mathfrak{g}, \, \mathfrak{g} ] $ (and $ {\rm span} (\bar{I}) $ with the center of $ \mathfrak{g}$)  and it is contained in $ \mathfrak{sl}(n, \, \mathbb{R} )\circledS \mathbb{R}^n $. 
A necessary condition for $ {\cal R}_G = G $ is that $ {\rm exp} \left( \left(\bar{{\cal L}}_{H_0} + \bar{\cal L}_D \right) t \right) \in {\rm exp}\left( \mathfrak{g}_1 \right) $ (i.e. to  $SL (n, \, \mathbb{R} ) $ or to $SL (n, \, \mathbb{R} )\circledS \mathbb{R}^n $) for some $t> 0 $ which is obviously never true.

For $ \bar{\cal L}_D$ unital, the reachable sets are balls in $ \mathbb{R}^n $ (centered in $0$) and are completely characterized by the following monotonicity property:
\begin{corollary}
If the system \eqref{eq:masterlin3} is accessible and if $ \bar{\cal L}_D$ unital then $ 
{\cal R}(\bar{\bm{\rho}}_i,  \, \leq T_1 ) \subsetneq {\cal R}(\bar{\bm{\rho}}_i,  \, \leq T_2 ) $ $ \forall \, 0 < T_1 < T_2 $ and $ {\cal R}(\bar{\bm{\rho}}_i) $ is the ball of radius $ \| \bar{\bm{\rho}}_i \| $. 
\end{corollary}
\proof
It follows from the observation above that $ \bar{\bm{\rho}}_\nu $ of norm $ \| \bar{\bm{\rho}}_\nu \| $ lies on the boundary of the set reachable from $ \bar{\bm{\rho}}_\nu $ by the integral curves of \eqref{eq:masterlin3}. If $ \bar{\bm{\rho}}_\nu =  \bar{\bm{\rho}} (T_1) = \Phi\left( {\cal T }{\rm exp} \int_0^{T_1} \bar{\cal L} (\tau) d \tau \right) (\bar{\bm{\rho}}_i ) $, then $ {\cal R}(\bar{\bm{\rho}}_i,  \, \leq T_2 ) =  {\cal R}(\bar{\bm{\rho}}_i,  \, \leq T_1 ) \cup  {\cal R}(\bar{\bm{\rho}}(T_1),  \, \leq T_2-T_1 ) $.
Notice that this does not require $ \bar{\cal L}_D $ to have a fixed point. Accessibility of \eqref{eq:masterlin3}, in fact, guarantees that $  \bar{\bm{\rho}} (t) $ can be placed on any point of the sphere of radius $ \| \bar{\bm{\rho}} (t)\| $ and therefore, as $ t\to \infty $, \eqref{eq:masterlin3} can be made to tend to the origin regardless of the existence of a fixed point for $ \bar{\cal L}_D $. Hence $ {\cal R}(\bar{\bm{\rho}}_i) $ is anything inside the ball of radius $ \| \bar{\bm{\rho}}_i \| $.
\qed

\section{Two level systems}
\label{sec:twol-control}
For two level systems, $ \bm{\rho}$ is the usual Bloch vector.
Call $ \lambda_k = \frac{1}{\sqrt{2}} \sigma_k $, $ k\in \{ 0, \, x,\, y, \, z \} $ the rescaled (identity and) Pauli matrices.
Then in the $\{ \lambda_k \} $ basis 
\beq
\rho = \begin{bmatrix}
\rho_{00} & \rho_{01} \\
\rho_{10} & \rho_{11}
\end{bmatrix} 
= \rho_0 \lambda_0 + \rho_x \lambda_x +\rho_y \lambda_y +\rho_z \lambda_z 
= \frac{1}{\sqrt{2}}
\begin{bmatrix}
\frac{1}{\sqrt{2}} +\rho_z & \rho_x - i \rho_y \\
\rho_x + i \rho_y & \frac{1}{\sqrt{2}} - \rho_z
\end{bmatrix} 
\label{eq:rho-2lev}
\eeq
and $ \bm{\rho} = [ \rho_x \; \rho_y \; \rho_z ]^T $, where $ \rho_k = \tr{\rho \lambda_k } $ i.e. $
\rho_0  =  \frac{1}{\sqrt{2}}$, $\rho_x =  \sqrt{2} \; {\rm Re}[\rho_{01}] $, $\rho_y  =  - \sqrt{2} \; {\rm Im}[\rho_{01}] $ and $ \rho_z  =  \frac{1}{\sqrt{2}} \left( \rho_{00} - \rho_{11} \right) $.
In our case, $ \{ \lambda_0 , \, \lambda_k \} = \sqrt{2} \lambda_k $, $ \{ \lambda_j , \, \lambda_k \} = \sqrt{2} \delta_{jk} \lambda_0 $, $ \forall \; j, k\in {\cal I} =\{ x,\, y, \, z \} $.
Similarly to \eqref{eq:rho-2lev}, the Hamiltonian $H$ can be written as 
\[
 H =\sum_{k\in {\cal I}} \sqrt{2} h_k \lambda_k 
=  \begin{bmatrix} h_z & h_x - i h_y \\
h_x + i h_y & - h_z 
\end{bmatrix}
\]
and, in the adjoint representation, from $- i {\rm ad}_H  =  \left(  - i {\rm ad}_H \right)_{pm} = \left( \sum_{l \in {\cal I}} h_l f_{lpm} \right)_{pm} $:
\beq
- i {\rm ad}_H =  
\begin{bmatrix} 
0 & -h_z & h_y \\
h_z & 0 & - h_x \\
- h_y & h_x & 0 
\end{bmatrix}
= h_x \begin{bmatrix} 
0 & 0 & 0 \\
0 & 0 & - 1 \\
0 & 1 & 0 
\end{bmatrix} 
+ h_y \begin{bmatrix} 
0 & 0 & 1 \\
0 & 0 & 0 \\
-1 & 0 & 0 
\end{bmatrix} 
+ h_z  \begin{bmatrix} 
0 & -1 & 0 \\
1 & 0 & 0 \\
0 & 0 & 0 
\end{bmatrix} 
\label{eq:coher-dyn}
\eeq
If $ \{ x, y, z \} = \{ 1, 2, 3\} $, specifying the coherent controls: $ h_k = \left( h_{0_k} + u_k \right)  $, $ k = 1, \, 2, \, 3 $, where $ h_{0_k} $ are the basis components of the time-independent free Hamiltonian $ H_0 $ and $ u_k = u_k (t) $, $ k = 1, \, 2, \, 3 $, the control parameters (some of the $ h_{0_k} $ or $ u_k $ may be $0$).
In the homogeneous coordinates, the vector field for the Hamiltonian acquires only a zero translation, and, from \eqref{eq:coher-dyn}, the infinitesimal generators of the coherent rotations are:
 \[
\bar{M}_{1} = \begin{bmatrix} 
0 & 0 & 0 & 0 \\
0 & 0 & 0 & 0 \\
0 & 0 & 0 & -1 \\
0 & 0 & 1 & 0 
\end{bmatrix} , \quad 
\bar{M}_{2} = \begin{bmatrix} 
0 & 0 & 0 & 0 \\
0 & 0 & 0 & 1 \\
0 & 0 & 0 & 0 \\
0 & -1 & 0 & 0 
\end{bmatrix} 
 , \quad 
\bar{M}_{3} = \begin{bmatrix} 
0 & 0 & 0 & 0 \\
0 & 0 & -1 & 0 \\
0 & 1 & 0 & 0 \\
0 & 0 & 0 & 0 
\end{bmatrix} 
\]
Hence $  \bar{{\cal L}}_{H_0 } = \sum_{k =1}^3 h_{0_k} \bar{M}_k $ and $ \bar{\cal L}_{H_k} = \bar{M}_k $, $ k=1,\, 2, \, 3$.
In this case \eqref{eq:ajkakj} simplifies to:
\beq
a_{jk} \bar{L}_{jk} + a_{kj} \bar{L}_{kj} = \begin{bmatrix}  0 & 0 \\
2 i a_{jk}^\Im \bm{v}_{jk} & (2 - \delta_{jk} )  a_{jk}^\Re L_{jk} 
\end{bmatrix}
\label{eq:ajk-2lev}
\eeq
The 9 degrees of freedom of $A$ (constrained by the positive semidefiniteness requirement) are captured by the 9 real parameters (reindexed cardinally)
\[
\left\{ a_4 , \, a_5 \, \ldots a_{12} \right\} = \left\{ a_{xy}^\Re , \, a_{xy}^\Im, \,  a_{xz}^\Re , \, a_{xz}^\Im, \,  a_{yz}^\Re , \, a_{yz}^\Im, \,   a_{xx}^\Re , \,  a_{yy}^\Re , \, a_{zz}^\Re  \right\} 
\]
In terms of $ a_4, \ldots a_{12}$, the matrix $A$ is:
\[
A = \begin{bmatrix}
a_{10} & a_4 + i a_5 & a_6 + i a_7 \\
a_4 - i a_5 & a_{11} & a_8 + i a_9 \\
 a_6 - i a_7 &  a_8 - i a_9 & a_{12} 
\end{bmatrix}
\]
In order to impose the positive semidefiniteness of $A$, a sufficient condition is that all the principal minors have nonnegative determinant, i.e. 
\beqa
& a_{10} \geq 0 , \qquad  a_{11} \geq 0 , \qquad  a_{12} \geq 0 & \label{eq:posA1}\\
& a_{10} a_{11} \geq a_4^2 + a_5 ^2 , \qquad a_{10} a_{12} \geq a_6^2 + a_7 ^2 , \qquad a_{11} a_{12} \geq a_8^2 + a_9 ^2 & \label{eq:posA2}\\
& a_{10} a_{11} a_{12} - a_{10} ( a_8^2 + a_9^2 ) - a_{11} ( a_6^2 + a_7^2 ) - a_{12} ( a_4^2+a_5^2 ) + 2a_4 ( a_6 a_8 + a_7 a_9 ) - 2 a_5 (a_6 a_9 - a_7 a_8) \geq 0
& \nonumber
\eeqa
The infinitesimal generators corresponding to this parameterization are linear combinations of the $ \bar{L}_{jk}$.
Numbering in the same fashion as the $ a_{jk}^\Re, a_{jk}^\Im $ parameters, we obtain the 9 linearly independent generators:
\[
\bar{M}_4=\bar{L}_{xy}+ \bar{L}_{yx} = \begin{bmatrix} 
0 & 0 & 0 & 0 \\
0 & 0 & 1 & 0 \\
0 & 1 & 0 & 0 \\
0 & 0 & 0 & 0 
\end{bmatrix} , \quad 
\bar{M}_5 =i \left( \bar{L}_{xy}-\bar{L}_{yx}\right) = \begin{bmatrix} 
0 & 0 & 0 & 0 \\
0 & 0 & 0 & 0 \\
0 & 0 & 0 & 0 \\
-2 & 0 & 0 & 0 
\end{bmatrix} 
\]
\[
\bar{M}_6 =\bar{L}_{xz}+\bar{L}_{zx} = \begin{bmatrix} 
0 & 0 & 0 & 0 \\
0 & 0 & 0 & 1 \\
0 & 0 & 0 & 0 \\
0 & 1 & 0 & 0 
\end{bmatrix}
 , \quad 
\bar{M}_7 =i \left( \bar{L}_{xz}-\bar{L}_{zx}\right) = \begin{bmatrix} 
0 & 0 & 0 & 0 \\
0 & 0 & 0 & 0 \\
2 & 0 & 0 & 0 \\
0 & 0 & 0 & 0 
\end{bmatrix} 
\]
\[
\bar{M}_8=\bar{L}_{yz}+\bar{L}_{zy} = \begin{bmatrix} 
0 & 0 & 0 & 0 \\
0 & 0 & 0 & 0 \\
0 & 0 & 0 & 1 \\
0 & 0 & 1 & 0 
\end{bmatrix} , \quad 
\bar{M}_9  = i \left( \bar{L}_{yz}-\bar{L}_{zy} \right) = \begin{bmatrix} 
0 & 0 & 0 & 0 \\
-2 & 0 & 0 & 0 \\
0 & 0 & 0 & 0 \\
0 & 0 & 0 & 0 
\end{bmatrix} 
\]
\[
\bar{M}_{ 10} = \bar{L}_{xx} = \begin{bmatrix} 
0 & 0 & 0 & 0 \\
0 & 0 & 0 & 0 \\
0 & 0 & -1 & 0 \\
0 & 0 & 0 & -1 
\end{bmatrix} , \; \;
\bar{M}_{ 11} =\bar{L}_{yy} = \begin{bmatrix} 
0 & 0 & 0 & 0 \\
0 & -1 & 0 & 0 \\
0 & 0 & 0 & 0 \\
0 & 0 & 0 & -1 
\end{bmatrix} 
 , \; \;
\bar{M}_{ 12} = \bar{L}_{zz} = \begin{bmatrix} 
0 & 0 & 0 & 0 \\
0 & -1 & 0 & 0 \\
0 & 0 & -1 & 0 \\
0 & 0 & 0 & 0 
\end{bmatrix} 
\]
The above expression of the matrix generators is very convenient for our purposes, because it splits the affine and linear parts of the action on $ \bm{\rho}$.
Furthermore, it makes it straightforward to check that $ {\rm Lie} \left( \bar{M}_1, \ldots, \bar{M}_{12} \right) = \mathfrak{gl}( 3 , \, \mathbb{R} )\circledS \mathbb{R}^3 $ (recall that $ {\rm dim}\: (\mathfrak{gl}( 3 , \, \mathbb{R} )\circledS \mathbb{R}^3 ) = 12 $).

In terms of the coherence vector and using homogeneous coordinates, eq. \eqref{eq:masterlin3} becomes:
\beq
\dot{\bar{\bm{\rho}}} = \sum_{k=1}^3 h_{0_k}  \bar{M}_k  \bar{\bm{\rho}} +  \sum_{k=1}^3 u_k \bar{M}_k  \bar{\bm{\rho}} + \sum_{k =4}^{12} a_{k} \bar{M}_{k} \bar{\bm{\rho}} , \qquad \bar{\bm{\rho}} \in \bar{\mathbb{B}}^3 
\label{eq:master-2lev2}
\eeq
Given $\bar{\cal L}_D $, the corresponding Lie algebra is
\[
 \mathfrak{g} = {\rm Lie} \left(  \sum_{k=1}^3 h_{0_k}  \bar{M}_k  + \sum_{k =4}^{12} a_{k} \bar{M}_{k} , \,  \sum_{k=1}^3 u_k \bar{M}_k \right) \subseteq\mathfrak{gl}( 3 , \, \mathbb{R} )\circledS \mathbb{R}^3 ;
\]
In general, $ \mathfrak{g} $ varies with the values of $ a_k $. A few prototypes of subalgebras obtained disregarding the assumption $A \geq 0 $ are reported in Table \ref{tab:subal-2lev}.
Once $ A\geq 0 $ is imposed, {\underline{\sf case 1}}, {\underline{\sf case 3}} and {\underline{\sf case 4}} are not anymore admissible (the argument is the same as in the proof of Theorem~\ref{thm:acc-contr-N}, Part 1).  
\begin{table}[ht]
\caption{Examples of subalgebras of $\mathfrak{gl}( 3 , \, \mathbb{R} )\circledS \mathbb{R}^3 $ obtained for different $A$ (not necessarily $ A\geq 0 $)}
\begin{center}
\begin{tabular}{|c|c|c|}
\hline
 & Coefficients $ a_{jk} $ & $ \mathfrak{g} $ \\
\hline
 $ \quad $ {\underline{\sf case 1}}  $ \quad $ & $ \quad $ $ a_5=a_7=a_9 =0 $; $ a_{10}, \, a_{11} , \, a_{12} $ s.t. $ {\rm tr}\left(  \sum_{k=10}^{12} a_k \bar{M}_k \right) = 0 $$ \quad $ &$ \quad $ $ \mathfrak{sl}(3, \, \mathbb{R} ) $$ \quad $ \\
\hline 
$ \quad $ {\underline{\sf case 2}}  $ \quad $ & $ a_5=a_7=a_9 =0 $; $ a_{10}, \, a_{11} , \, a_{12} $ s.t. $ {\rm tr}\left( \sum_{k=10}^{12} a_k \bar{M}_k  \right) \neq 0 $ & $ \mathfrak{gl}(3, \, \mathbb{R} ) $\\
\hline 
$ \quad $ {\underline{\sf case 3}}  $ \quad $ & $ a_4=a_6=a_8 = a_{10}= a_{11} = a_{12} =0 $ & $ {\rm ad}_{\mathfrak{su}(2)} \circledS \mathbb{R}^3  $\\ 
\hline
$ \quad $ {\underline{\sf case 4}}  $ \quad $ & $ a_{10}, \, a_{11} , \, a_{12} $ s.t. $ {\rm tr}\left( \sum_{k=10}^{12} a_k \bar{M}_k  \right) = 0 $ & $ \mathfrak{sl}(3, \, \mathbb{R} )\circledS \mathbb{R}^3 $\\
\hline
$ \quad $ {\underline{\sf case 5}}  $ \quad $ & $ a_{10}, \, a_{11} , \, a_{12} $ s.t. $ {\rm tr}\left( \sum_{k=10}^{12} a_k \bar{M}_k \right)  \neq 0 $ &$ \quad $  $ \mathfrak{gl}(3, \, \mathbb{R} )\circledS \mathbb{R}^3 $$ \quad $ \\
\hline
$ \quad $ {\underline{\sf case 6}}  $ \quad $ & $ a_4= \ldots = a_9 =0$,  $ \; a_{10} = a_{11} = a_{12} $ &$ \quad $ $ {\rm ad}_{\mathfrak{su}(2)}  \oplus {\rm span} (\bar{I}) $ $ \quad $ \\
\hline
$ \quad $ {\underline{\sf case 7}}  $ \quad $ & $ a_4= a_6 = a_8 =0$,  $ \; a_{10} = a_{11} = a_{12} $ &$ \quad $ $ ( {\rm ad}_{\mathfrak{su}(2)}  \oplus {\rm span}(\bar{I})) \circledS \mathbb{R}^3 $ $ \quad $ \\
\hline
\end{tabular}
\label{tab:subal-2lev}.
\end{center}
\end{table}

The two level version of Theorem~\ref{thm:acc-contr-N} is then:
\begin{theorem}
\label{thm-contr-2lev}
For a two-level master equation, under Assumptions {\bf A1} and {\bf A2} we have:
\begin{enumerate}
\item the system \eqref{eq:master-2lev2} is accessible in $ \bar{\mathbb{B}} ^3 $ for $ \bar{\cal L}_D \neq \alpha \bar{I} + \sum_{k=5,7,9} a_k \bar{M}_k $, $\alpha\in \mathbb{R} , \, \alpha< 0 $.
\item the system \eqref{eq:master-2lev2} is never small-time nor finite-time controllable in $ \bar{\mathbb{B}} ^3 $ for $ {\cal L}_D \neq 0 $;
\end{enumerate}
\end{theorem}

\proof
The assumption $ \bar{\cal L}_D \neq \alpha \bar{I} + \sum_{k=5,7,9} a_k \bar{M}_k $, $ \alpha < 0 $, rules out {\underline{\sf case 6}} and  {\underline{\sf case 7}} of Table \ref{tab:subal-2lev}.
By exclusion, or by exhaustive computation using the structure constant of Appendix~\ref{app:struc-const}, any nonnull $ \bar{\cal L}_D $ such that $ \bar{\cal L}_D \neq \alpha \bar{I} + \sum_{k=5,7,9} a_k \bar{M}_k $ generates the Lie algebra of {\underline{\sf case 2}} or {\underline{\sf case 5}} as required by Theorem~\ref{thm:acc-contr-N}.  
\qed

\subsection{Examples}
\label{subs:examples}

In quantum information processing, some of the $ \bar{M}_k $ admit well-known physical interpretations in terms of nonunitary quantum operations on a qubit normally used in the theory of error correction.
For example $ \bar{M}_{10} $, $ \bar{M}_{11} $ and $ \bar{M}_{12} $ are, respectively, the infinitesimal generators of the one-parameter semigroups corresponding to {\em bit flip}, {\em bit-phase flip} and {\em phase flip} channels (see \cite{Nielsen1}, Sec.~8.4) and so a {\em depolarizing channel} has $ a_{10} $, $ a_{11} $ and $ a_{12} $ all nonnull and equal.

\subsubsection{Depolarizing channel ($ \bar{\cal L}_D $ unital)}
The depolarizing channel is given by $ \bar{\cal L}_D = \alpha \bar{I} $, $ \alpha < 0 $.
Since $ \bar{\cal L}_D $ commutes with everything, in this case the system is not accessible and furthermore its integral curves are not at all modified by coherent control. They will be pointing ``isotropically'' to the origin in $ \mathbb{R}^3 $.

\subsubsection{Phase flip ($ \bar{\cal L}_D $ unital)}
The phase flip channel is also known as phase damping or pure coherence channel and it is given by $ \bar{\cal L}_D $ aligned with $ \bar{M}_{12}$.
The effect of this one-parameter semigroup is to ``contract'' the Block sphere along the $\lambda_x$ and $\lambda_y$ directions, leaving it untouched along $ \lambda_z $. 
As an example, check the accessibility property in correspondence of the following simple master equation:
\[
\dot{\bar{\bm{\rho}}} = \left( u_1 \bar{M}_1 + u_2 \bar{M}_2 + u_3 \bar{M}_3 + a_{12} \bar{M}_{12} \right)\bar{\bm{\rho}} 
\]
i.e. controls available along all the three directions and no free Hamiltonian.
The Lie algebra $ {\rm Lie} \left\{  \bar{M}_1, \, \bar{M}_2, \, \bar{M}_3,\,  \bar{M}_{12} \right\} $ is computed using the structure constants given in Appendix \ref{app:struc-const} and the Jacobi identity to eliminate terms not linearly independent.
\begin{itemize}
\item first level Lie brackets
\[
[ \bar{M}_1, \, \bar{M}_{12} ]= -\bar{M}_8 , \qquad 
[ \bar{M}_2, \, \bar{M}_{12} ]=  \bar{M}_6 
\]
\item second level Lie brackets
\[
[ \bar{M}_1, \, \bar{M}_{6} ]= -\bar{M}_4 , \qquad 
[ \bar{M}_1, \, \bar{M}_{8} ]=   \bar{M}_{12} - \bar{M}_{11} , \qquad
[ \bar{M}_2, \, \bar{M}_{6} ]=   \bar{M}_{10} - \bar{M}_{11} 
\]
\end{itemize}
Therefore
\[
 \mathfrak{g}  =  \left\{  \bar{M}_1, \, \bar{M}_2, \, \bar{M}_3,\, 
  \bar{M}_{4}  , \,   \bar{M}_{6}, \, \bar{M}_{8}, \,  \bar{M}_{10} - \bar{M}_{12}, \,   \bar{M}_{12} - \bar{M}_{11}     , \,  \bar{M}_{12} \right\} \\
=  \mathfrak{gl}(3, \mathbb{R})
\]
and the process is accessible.
Notice that $ \bar{M}_{10} - \bar{M}_{12}$ and $ \bar{M}_{12} - \bar{M}_{11} $ are {\em traceless} i.e. they belong to $ \mathfrak{sl}(3, \mathbb{R})$ (unlike $ \bar{M}_{12} $) and therefore that $ \mathfrak{g}_1 = \mathfrak{sl}(3, \mathbb{R})$, as expected.

In this case, as it is easy to check (see also \cite{Alicki1}, Part 2, Sec. II.5) $ \bar{\cal L}_D $ is not uniquely relaxing, i.e. a fixed point for the uncontrolled system does not exist. 
Thus the asymptotic value depends from the initial condition and $ \lim_{t\to \infty} \bar{\bm{\rho}} = [ \rho_0 \, 0 \, 0 \, \rho_z (0) ]^T $. 
Once the control is added, however, the controlled system can be made to converge to any $ \tilde{\rho}_z  $ in the interval $ [ - \rho_z (0) , \, \rho_z (0) ]$.
In particular, if $  \tilde{\rho}_z =0 $ then $ {\cal R}(\bar{\bm{\rho}}(0) ) = \{ \bar{\bm{\rho}} \in \bar{\mathbb{B}}^n \; \text{ s.t. } \; \|  \bar{\bm{\rho}} \| \leq \|  \bar{\bm{\rho}} (0) \| \} $.

A Lie algebraic method {\em per se} is normally not constructive. 
However, what it tells in this case is that full accessibility is achieved only at the second level of brackets.
Therefore a series expansion cannot be truncated before that, if one wants to assure the generation of group actions in arbitrary directions.

To understand what is happening to the integral curves of the system, it is convenient to split $ \bar{\cal L}_D $ into part in $\mathfrak{sl}(3) $ and part in $ \mathfrak{gl}(3) \setminus \mathfrak{sl}(3) $ as in the proof of Theorem~\ref{thm:acc-contr-N}: $  \bar{\cal L}_D = \alpha \bar{I} + \tilde{\cal L}_D $, $ \alpha < 0 $, $  \tilde{\cal L}_D \in \mathfrak{sl}(3)$.
For $  \bar{\cal L}_D = a_{12} \bar{M}_{12} $
\[
 \bar{\cal L}_D = a_{12} \bar{M}_{12} =\alpha \bar{I} +  \tilde{\cal L}_D  = -\frac{2 a_{12} }{3} \bar{I} + \frac{a_{12}}{3} \begin{bmatrix} 0 & 0 & 0 & 0 \\
0 & -1 & 0  & 0 \\ 0& 0 & -1 & 0 \\ 0 & 0&0 & 2 \end{bmatrix} 
\]
If $F(t) = \tilde{\cal L}_D + \sum_{k=1}^3 ( h_{0_k} + u_k (t) )\bar{M}_{k} $, since $ [ \bar{I}, \, F ] = 0 $ $ \forall F \in \mathfrak{gl}(3, \, \mathbb{R}) $, the flow of the system can be written as the exponential 
\[
g(t) = {\cal T} {\rm exp}\int_{0}^t  ( \alpha \bar{I} + F(\tau) ) d \tau = {\rm exp} ( t\alpha \bar{I} )\, {\cal T} {\rm exp}\int_{0}^t  F(\tau) d \tau 
\]
and its action on $ \bar{\bm{\rho}} $ as $\bar{\bm{\rho}}(t) = {\rm exp} ( t\alpha \bar{I} ){\cal T} {\rm exp}\int_{0}^t  F(\tau) d \tau  \bar{\bm{\rho}}(0) $.
The ``isotropic'' contraction $ {\rm exp}  (t \alpha \bar{I}) $ corresponds to the depolarizing channel $ \alpha/2 ( \bar{M}_{10} +   \bar{M}_{11} + \bar{M}_{12} )$ and cannot be reversed. 
Furthermore, the complete positivity constraint $ A\geq 0 $ imposes that the $ \alpha \bar{I} $ part must be dominant with respect to the $ F(t) $ part.
Thus, regardless of the control action, the overall result is a contraction of the flow in $ \bar{\mathbb{B}}^3$.

\subsubsection{Amplitude damping ($ \bar{\cal L}_D $ affine)}
\label{sbus:spont-em}
In terms of the master equation, the amplitude damping channel corresponds to an atomic system with spontaneous emission.
In a two-level system, the excited state $ \ket{1}$ can decay to ground state $ \ket{0}$ while emitting a photon.
The process of spontaneous emission is characterized in terms of the atomic ladder operators $ \sigma_\pm = \sigma_x \pm i \sigma_y $ and of the damping coefficient $\gamma_\downarrow  $ ($\gamma_\downarrow > 0 $) as (see ex. the survey \cite{Alicki2} and \cite{Nielsen1}, Sec. 8.4.1)
\beq
\dot{\rho} = -i {\rm ad}_H (\rho) -\frac{\gamma_\downarrow }{2} \left( 2 \sigma_- \rho \sigma_+ - \sigma_+ \sigma_- \rho - \rho \sigma_+ \sigma_- \right)
\label{eq:sp-em}
\eeq
If, for example, $ H = \sqrt{2} \left( h_{0_3} \lambda_3 + \sum_{k=1}^3 u_k \lambda_k \right) $ then
\beqa
\dot{\bar{\bm{\rho}}} 
& = & \begin{bmatrix} 
0 & 0 & 0 & 0 \\
0 & 0 & -h_{0_3} - u_3 & u_2 \\
0 & h_{0_3} + u_3 & 0 & - u_1 \\
0 & -u_2 & u_1 & 0 
\end{bmatrix} \bar{\bm{\rho}} +  
 \gamma_\downarrow \begin{bmatrix} 
0 & 0 & 0 & 0 \\
0 & -\frac{1}{2}  & 0 & 0  \\
0 & 0 & -\frac{1}{2} & 0 \\
1 & 0 & 0 & -1 
\end{bmatrix} \bar{\bm{\rho}} \nonumber \\
& = & h_{0_3} \bar{M}_3 \bar{\bm{\rho}} +  \sum_{k=1}^3 u_k \bar{M}_k  \bar{\bm{\rho}} + \frac{\gamma_\downarrow}{2} \left( \bar{M}_{10} +  \bar{M}_{11} - \bar{M}_{5} \right)  \bar{\bm{\rho}}
\label{eq:sp-em2}
\eeqa
Since the unital part of $ \bar{\cal L}_D $ is not proportional to the identity, by Theorem~\ref{thm-contr-2lev} spontaneous emission is an accessible process.
From \eqref{eq:ajk-2lev}, the relaxation matrix is 
\[
A = \frac{\gamma_\downarrow}{2} \begin{bmatrix} 1 & -i & 0 \\ i & 1 & 0 \\ 0 & 0& 0 
\end{bmatrix}
\] 
As the unital part of $ \bar{\cal L}_D $ is invertible, \eqref{eq:sp-em2} has a fixed point. Since $ A\geq 0 $ but not $ A > 0 $, $ \bar{\cal L}_D $ lies on an exposed face of the cone $ A \geq 0 $ and in correspondence the fixed point lies on $ \partial \bar{\mathbb{B}}^3 $, i.e. it is a pure state.
Thus, asymptotically, $ \bar{\cal L}_D $ admits a reachable set such that $ {\rm cl} ( {\cal R}(\bar{\bm{\rho}}(0) ) ) = \bar{\mathbb{B}}^3$.
As mentioned in the proof of Theorem~\ref{thm:acc-contr-N}, coherent control can speed up the ``purification'' of $\bar{\bm{\rho}} $ only for certain values of the initial condition. 
In fact, from 
\[
\frac{1}{2} \frac{d \| \bar{\bm{\rho}} \|^2}{d t} = \llangle \dot{  \bar{\bm{\rho}}} , \,  \bar{\bm{\rho}} \rrangle =\llangle  h_{0_3} \bar{M}_3 \bar{\bm{\rho}} , \,  \bar{\bm{\rho}} \rrangle + \llangle  \sum_{k=1}^3 u_k \bar{M}_k  \bar{\bm{\rho}} , \,  \bar{\bm{\rho}} \rrangle + \frac{\gamma_\downarrow}{2} \llangle \left( \bar{M}_{10} +  \bar{M}_{11} - \bar{M}_{5} \right)  \bar{\bm{\rho}}   , \,  \bar{\bm{\rho}} \rrangle 
\]
only the last term gives a nontrivial contribution and can become positive for example in correspondence of $ \rho_z $ positive and $ \rho_x $, $ \rho_y $ small enough:
\[
 \frac{\gamma_\downarrow}{2} \llangle \left( \bar{M}_{10} +  \bar{M}_{11} - \bar{M}_{5} \right)  \bar{\bm{\rho}}  , \,  \bar{\bm{\rho}} \rrangle = \frac{\gamma_\downarrow}{2} \left( \rho_0 \rho_z - \frac{1}{2} \left( \rho_x^2 + \rho_y ^2 + \rho_z ^2 \right) \right) =   \frac{\gamma_\downarrow}{2} \left( \rho_0 \rho_z -  \frac{1}{2} \| \bm{\rho} \| ^2 \right)
\]
However, the purification process remains an asymptotic process since as soon as coherent control has brought $ \bm{\rho}$ to $ [ 0 \, 0\, + |\rho_z | ] $, then purification can occur only because of $ \bar{\cal L}_D $. 
Once again, notice how the role of $ \bar{\cal L}_D $ is essential in move around in the reachable set.

\section{Conclusion}
The aim of this work is to shed light on the possibility and limits of coherent control of Markovian master equations using standard tools from geometric control theory.
It turns out that there is a remarkable complementarity between admissible quantum dynamical semigroups and controllability: in the vector of coherences representation an admissible $ \bar{\cal L}_D $ has to have a nonull component along the nonzero-trace one-dimensional vector subspace of the Lie algebra of $n\times n $ matrices (or its semidirect extensions). A component in that direction guarantees noncontrollability in small and finite time.
In the simple cases of unital dissipation operators, the fact that the ``uncontrollable'' direction has dimension 1 allows to obtain an order relation among the sets reachable at different times by means of arbitrary coherent controls.

\section{Acknowledgments}
The author wishes to thank A. A. Agrachev, A. J. Landahl and S. Lloyd for helpful comments and discussions.

\appendix

\section{Structure constants of $ \mathfrak{gl}(3 , \mathbb{R} )\circledS \mathbb{R}^3  $}
\label{app:struc-const}
For the real Lie algebra $ \mathfrak{gl}(3 , \mathbb{R} ) $ in the basis $ \bar{M}_1 , \ldots , \bar{M}_{12} $, the structure constants, call them $ c_{jk}^l$, are real but not totally skew-symmetric:
\[
\begin{split}
c_{1,2}^3 =c_{1,4}^6 =c_{1,5}^7=c_{1,8}^{12}=c_{1,11}^8=c_{2,3}^1=c_{2,6}^{10}=c_{2,8}^4=c_{2,12}^6=c_{3,4}^{11}=c_{3,6}^8=c_{3,7}^9=c_{3,10}^4=c_{4,8}^2=1 \\
c_{4,10}^3 =c_{5,8}^7=c_{5,10}^5=c_{5,11}^5=c_{6,9}^{5}=c_{6,12}^{2}=c_{7,8}^{5}=c_{7,10}^{7}=c_{7,12}^{7}=c_{8,11}^{1}=c_{9,11}^{9}=c_{9,12}^{9}=1 \\
c_{1,3}^{2}=c_{1,6}^{4}=c_{1,7}^{5}=c_{1,8}^{11}=c_{1,12}^{8}=c_{2,4}^{8}=c_{2,5}^{9}=c_{2,6}^{11}=c_{2,9}^{5}=c_{2,10}^{6}=c_{3,4}^{10}=c_{3,8}^{6}=-1 \\
c_{3,9}^{7}=c_{3,11}^{4}=c_{4,6}^{1}=c_{4,7}^{9}=c_{4,9}^{7}=c_{4,11}^{3}=c_{5,6}^{9}=c_{6,8}^{3}=c_{6,10}^{2}=c_{8,12}^{1}=-1
\end{split}
\]

\bibliographystyle{abbrv}

\end{document}